\documentclass[%
reprint,
prd,
amsmath,amssymb,aps,
]{revtex4-2}
\usepackage [latin1]{inputenc}
\usepackage{graphicx}
\usepackage{dcolumn}
\usepackage{bm}

\usepackage{ifthen} 
\usepackage{fp,calc}
\usepackage{lastpage}
\IfPackageLoadedTF{color}{}{\usepackage{color}}

\usepackage{fancyhdr}
\usepackage{adjustbox}
\usepackage[english]{babel}

\IfPackageLoadedTF{xcolor}{}{\usepackage[svgnames, table]{xcolor}}
\usepackage{layout}
\usepackage{enumitem}
\usepackage[normalem]{ulem}
\usepackage{parskip}
\usepackage{multirow}
\usepackage{comment}
\usepackage[flushleft]{threeparttablex}

\newboolean{articletitles}
\setboolean{articletitles}{true} 
\newboolean{allauthors}
\setboolean{allauthors}{false} 

\providecommand\mnras{Mon.\ Not.\ Roy.\ Astron.\ Soc. }
\providecommand\aap{Astron.\ Astrophys. }
\providecommand\prd{Phys.\ Rev.\ D}

\providecommand\jcap{JCAP }

\providecommand\apj{Astrophys.\ J. }
\providecommand\aj{Astron.\ J. }

\newcommand{\hJ}{\hat{J}}
\newcommand{\hf}{\hat{f}}
\newcommand{\HH}{\mathcal{H}}

\begin{document}

\title{Testing General Relativity through the $E_G$ Statistic \\ Using the Weyl Potential and Galaxy Velocities}

\author{Nastassia Grimm}
\email[]{nastassia.grimm@unige.ch}
\affiliation{D\'epartement de Physique Th\'eorique and Center for Astroparticle Physics, Universit\'e de Gen\`eve, Quai E. Ansermet 24, CH-1211 Geneve 4, Switzerland}

\author{Camille Bonvin}
\email[]{camille.bonvin@unige.ch}
\affiliation{D\'epartement de Physique Th\'eorique and Center for Astroparticle Physics, Universit\'e de Gen\`eve, Quai E. Ansermet 24, CH-1211 Geneve 4, Switzerland}

\author{Isaac Tutusaus}
\email[]{isaac.tutusaus@irap.omp.eu}
\affiliation{Institut de Recherche en Astrophysique et Plan\'etologie (IRAP), Universit\'e de Toulouse, CNRS, UPS, CNES, 14 Av.~Edouard Belin, 31400 Toulouse, France}

\date{\today}

\begin{abstract}
We combine measurements of galaxy velocities from galaxy surveys with measurements of the Weyl potential from the Dark Energy Survey to test the consistency of General Relativity at cosmological scales. Taking the ratio of two model-independent observables -- the growth rate of structure and the Weyl potential --  we obtain new measurements of the $E_G$ statistic with precision of $6.0-11.3\%$ at four different redshifts. These measurements provide a considerable improvement to past measurements of $E_G$. They confirm the validity of General Relativity at four redshifts, with a deviation of at most $1.6\sigma$ from the predicted values. Contrary to conventional methods that rely on a common galaxy sample with spectroscopic resolution to measure two types of correlations, we directly combine two observables that are independent of the galaxy bias. This provides a novel approach to testing the relation between the geometry of our Universe and the motion of galaxies with improved precision.
\end{abstract}

\maketitle


\section{Introduction}

The observation of the accelerated expansion of our Universe in 1998~\cite{SupernovaSearchTeam:1998fmf,Perlmutter_1999} has called our understanding of the cosmos into question. This unexpected behaviour can either be explained by the presence of a new form of energy with negative pressure, called dark energy; or by changing the laws of gravity at cosmological scales. To determine which of these paradigms is correct, it is necessary to test the theory of gravity. Gravity determines, on one hand, how space and time are bent by light and matter; and, on the other hand, it governs how light and matter move in this distorted geometry. In General Relativity (GR) these two concepts are directly linked by Einstein's equations and conservation equations. If gravity is modified, however, the relation between the content of the Universe and its geometry typically changes, see e.g., Ref.~\cite{Clifton:2011jh} for a review.

Given the complexity of the landscape of theories beyond GR, it is unfeasible to test and constrain all theories beyond GR one by one. A powerful approach consists instead in building model-independent tests that probe particular features of GR and would immediately signal any deviation in the data. One example is the so-called $E_G$ statistic proposed in 2007 by Zhang et al.~\cite{Zhang:2007nk}. The idea of this test is to compare the distortion of space-time with the velocities of astrophysical objects, for example galaxies, to determine if their relation obeys GR or not. More precisely this test measures the ratio of the Weyl potential, i.e., the sum of the two gravitational potentials $\Psi_W=(\Phi+\Psi)/2$, and the galaxy peculiar velocity $V$. In GR, this ratio takes a specific value, while a modified theory of gravity generally yields a different prediction. Hence, any deviation from the expected $E_G$ value would indicate a departure from GR. The strength of this test lies in the fact that it does not require any modelling of theories beyond GR: one simply compares two observable quantities to test the consistency of GR. Given the recent tensions with respect to the $\Lambda$CDM model of cosmology, e.g.\ the $H_0$ tension~\cite{Verde:2019ivm,DiValentino:2021izs}, the kinematic dipole tension~\cite{Secrest:2020has,Dalang:2021ruy,Secrest:2022uvx,Wagenveld:2023kvi} or the $\sigma_8$-tension~\cite{DES:2021wwk,Garcia-Garcia:2021unp,Abdalla:2022yfr,DES:2022xxr,Busch:2022pcx}, as well as the recent baryon acoustic oscillation measurements by the Dark Energy Spectroscopic Instrument
suggesting an evolution of the dark energy component when combined with supernovae measurements~\cite{DESI:2024mwx}, the $E_G$ statistic provides a clean and robust way of assessing the validity of GR.

Observationally, we cannot access the Weyl potential and galaxy velocities directly. Instead, galaxy surveys measure correlation functions involving these quantities, averaged over the sky. The current method to constrain $E_G$ therefore consists in measuring correlations of the Weyl potential, as well as of the galaxy velocities, with a common galaxy sample. Then, by taking the ratio between these two sets of correlation functions, the contribution from galaxy clustering cancels out which leads to the desired $E_G$ quantity. In practice, the situation is slightly more complex, since direct measurements of galaxy velocities only reach up to $z\simeq 0.02$, see e.g., Refs.~\cite{Howlett:2017asq, Huterer:2016uyq, Hudson:2012gt, Turnbull:2011ty, Davis:2010sw, Song:2010rm}. Velocity-density correlations are therefore reconstructed from the product of galaxy clustering and the quantity $\beta\equiv f/b$, which can be directly measured from redshift-space distortions~\cite{Kaiser:1987qv,Hamilton:1995px} (here $f$ is the growth rate and $b$ the galaxy bias). More precisely, $E_G$ can be obtained via the ratio,
\begin{equation}
    E_G=\Gamma\frac{C_l^{\kappa g}}{\beta C_l^{gg}}\,,
\end{equation}
where $C_l^{gg}$ is the galaxy clustering angular power spectrum, and $C_l^{\kappa g}$ denotes the cross-correlation of galaxy clustering with gravitational lensing. $\Gamma$ is a normalisation pre-factor, see e.g., Ref.~\cite{Pullen2016} for detail.

This technique has been successfully used with various data sets~\cite{Reyes:2010tr, Blake:2016fcm, Amon2018MNRAS.479.3422A, Alam:2016qcl, Pullen2016, Wenzl:2024sug}, providing measurements of $E_G$ over a wide range of redshifts. While some of the measurements are in perfect agreement with the predictions from GR~\cite{Reyes:2010tr, Blake:2016fcm, Amon2018MNRAS.479.3422A, Alam:2016qcl, Wenzl:2024sug}, others reveal some tensions~\cite{Blake:2016fcm, Amon2018MNRAS.479.3422A, Pullen2016}. Recently, Ref.~\cite{Abidi_2023_10419} showed that 21\,cm intensity mapping will provide a robust way to measure $E_G$ with the coming generation of surveys. One drawback of the $E_G$ statistic is that it requires a common galaxy sample to measure $C_\ell^{\kappa g}, C_\ell^{gg}$ and $\beta$, in order to cancel the galaxy bias. Since $\beta$ can only be precisely measured from spectroscopic surveys, this means that one cannot use the statistical power of photometric samples (that contain typically many 
more galaxies than spectroscopic samples) to measure $C_\ell^{\kappa g}$ and $C_\ell^{gg}$. As a consequence, the precision in the measurement of $E_G$ is degraded. This is quite common to model-independent tests, which are usually less constraining than direct analyses of specific theories beyond GR, due to additional requirements that are necessary to ensure the model independence.

In this letter, we present a novel method to measure $E_G$, which is not affected by the degradation inherent to standard methods. We combine 22 measurements of the growth rate of structure, $f\sigma_8$ ($\sigma_8$ denotes the clustering amplitude), from various spectroscopic surveys~\cite{Howlett:2017asq, Huterer:2016uyq, Hudson:2012gt, Turnbull:2011ty, Davis:2010sw, Song:2008qt, Blake:2013nif, eBOSS:2020yzd, Blake:2012pj, Pezzotta:2016gbo, Okumura:2015lvp, eBOSS:2018yfg}, with novel measurements of the Weyl potential from the photometric Dark Energy Survey (DES)~\cite{Tutusaus:2023aux}. Hence, our measurements of $E_G$ do not rely on a common spectroscopic data set. Instead, we use model-independent measurements of the evolution of galaxy velocities (encoded in $f\sigma_8$) and the evolution of the Weyl potential (encoded into a function $\hJ$, called Weyl evolution), and combine them to obtain $E_G$.  This allows us to benefit, at the same time, from the full precision of spectroscopic samples to measure velocities and from the full statistical power of photometric samples to measure the Weyl potential, which significantly improves the precision of the $E_G$ measurements. We obtain values of $E_G$ at four redshifts, with precision of $6.0-11.3\%$, providing a considerable improvement to previous measurements with $13.3-30.2\%$ precision. We find that $E_G$ is fully compatible with GR at the $1\sigma$ level in three of the four redshift bins, and at the $1.6\sigma$ level in the second bin. 

\section{Methodology}

Galaxy surveys measure the galaxy clustering correlation function in redshift-space. Since galaxy peculiar velocities break statistical isotropy, the correlation function can be decomposed into multipoles, namely a monopole, quadrupole, and hexadecapole. By combining measurements of these different multipoles, one can measure the quantity $\hf(z)\equiv f(z)\sigma_8(z)$ at redshift $z$ in a model-independent way, i.e., without assuming a specific theory of gravity. Here, $f(z)=\mathrm d\ln \delta/\mathrm d \ln a$ is the growth rate of structure, which encodes the evolution of the galaxy peculiar velocities, and $\sigma_8(z)$ is the clustering amplitude in spheres of 8\,Mpc/$h$. 

Recently it was shown that weak lensing data can be used in a similar spirit, to obtain model-independent measurements of the evolution of the Weyl potential~\cite{Tutusaus:2023aux}. The evolution of the Weyl potential with redshift can be encoded in a function $\hJ$, defined through~\cite{Tutusaus:2022cab}
\begin{align}
\Psi_W(k,z)= \left(\frac{\HH(z)}{\HH(z_*)} \right)^2\hJ(z)\, \sqrt{\frac{B(k,z)}{B(k,z_*)}}\frac{\Psi_W(k,z_*)}{\sigma_8(z_*)}\, . 
\label{eq:hatJ}
\end{align}
Here, $\HH$ is the Hubble parameter in conformal time, and $z_*$ is a reference redshift chosen to be well in the matter-dominated era, before acceleration started. Moreover, the boost factor $B(k,z)$ encodes the non-linear evolution of matter density perturbations at small scales. As shown in Ref.~\cite{Tutusaus:2023aux}, $\hJ$ can be directly measured by combining galaxy-galaxy lensing correlations with galaxy clustering correlations. These measurements are model-independent, i.e., they do not assume any theory of gravity. Similarly to the measurements of $\hf$ they rely however on one assumption: that at $z_*$, before the phase of accelerated expansion, GR is recovered. This is a reasonable assumption, given that the main motivation to search for theories beyond GR is the late-time accelerated expansion of the Universe, while early-time cosmological observations are compatible with GR. As constraints from the Cosmic Microwave Background~\cite{Planck:2015bue} and from nucleosynthesis~\cite{Asimakis:2021yct} largely restrict the scope of allowed theories, it is a common approach to assume that deviations from GR only occur at late time~\cite{Ferreira:2019xrr}, driving the accelerated expansion of the Universe. 

Given measurements of $\hf$ and $\hJ$, $E_G$ can then be obtained by taking the ratio, 
\begin{align}
\label{eq:EG}
E_G(z)\equiv\left(\frac{\HH(z)}{\HH_0}\right)^2\frac{1}{1+z}\,\frac{\hJ(z)}{\hf(z)}\, . 
\end{align}
Note that here we have normalised the ratio by the Hubble parameter, to recover the standard expression for $E_G$ defined in Ref.~\cite{Zhang:2007nk}, such that we can compare with the values previously measured in the literature. 
In GR, using that $\Phi=\Psi$ and that $\Phi$ is related to the density via the Poisson equation, $\hJ$ becomes $\hJ(z)=\Omega_{\rm m}(z)\sigma_8(z)$, where $\Omega_{\rm m}(z)$ is the matter density parameter at redshift $z$, and $E_G$ reduces to $E_G=\Omega_{\rm m,0}/f(z)$. In modified theories of gravity, however, $\Phi$ usually differs from $\Psi$ and the Poisson equation is typically modified, leading to different values of $E_G$. Note that without the normalisation in front of the ratio $\hJ/\hf$ in Eq.~\eqref{eq:EG}, we would instead obtain $E_G=\Omega_{\rm m}(z)/f(z)$ in GR. 

\begin{table}
\centering
\caption{List of $\hat{f}(z)$ measurements used in this work.} \label{tab:fhat_val}
\begin{tabular}{c c c c} 
\hline 
\multicolumn{4}{c}{\vspace{-8pt}} \\
Dataset & $z$ & $\hat{f}(z)$ & Ref. \\
\multicolumn{4}{c}{\vspace{-8pt}} \\
\hline\hline 
\multicolumn{4}{c}{\vspace{-8pt}} \\
2MTF & $0.001$ & $0.505 \pm 0.085$ & \cite{Howlett:2017asq} \\
6dFGS+SNIa & $0.02$ & $0.4280 \pm 0.0465$ & \cite{Huterer:2016uyq}\\
IRAS+SNIa & $0.02$ & $0.398 \pm 0.065$ & \cite{Hudson:2012gt, Turnbull:2011ty}\\
2MASS & $0.02$ & $0.314 \pm 0.048$ & \cite{Hudson:2012gt, Davis:2010sw}\\
2dFGRS & $0.17$ & $0.510 \pm 0.060$ & \cite{Song:2008qt}\\
GAMA & $0.18$ & $0.360 \pm 0.090$ & 
\cite{Blake:2013nif}\\
GAMA & $0.38$ & $0.440 \pm 0.060$ & 
\cite{Blake:2013nif}\\
SDSS-IV (MGS) & $0.15$ & $0.53 \pm 0.16$ & \cite{eBOSS:2020yzd} \\
SDSS-IV (BOSS Galaxy) & $0.38$ & $0.497 \pm 0.045$ & \cite{eBOSS:2020yzd}\\
SDSS-IV (BOSS Galaxy) & $0.51$ & $0.459 \pm 0.038$ & \cite{eBOSS:2020yzd}\\
SDSS-IV (eBOSS LRG) & $0.70$ & $0.473 \pm 0.041$ & \cite{eBOSS:2020yzd}\\
SDSS-IV (eBOSS ELG) & $0.85$ & $0.315 \pm 0.095$ & \cite{eBOSS:2020yzd}\\
WiggleZ & $0.44$ & $0.413 \pm 0.080$ & \cite{Blake:2012pj} \\
WiggleZ & $0.60$ & $0.390 \pm 0.063$ & \cite{Blake:2012pj}\\
WiggleZ & $0.73$ & $0.437 \pm 0.072$ & \cite{Blake:2012pj}\\
Vipers PDR-2 & $0.60$ & $0.550 \pm 0.120$ & \cite{Pezzotta:2016gbo} \\
Vipers PDR-2 & $0.86$ & $0.400 \pm 0.110$ & \cite{Pezzotta:2016gbo} \\
FastSound & $1.40$ & $0.482 \pm 0.116$ & \cite{Okumura:2015lvp} \\
SDSS-IV (eBOSS Quasar) & $0.978$ & $0.379 \pm 0.176$ & \cite{eBOSS:2018yfg}\\
SDSS-IV (eBOSS Quasar) & $1.230$ & $0.385 \pm 0.099$ & \cite{eBOSS:2018yfg}\\
SDSS-IV (eBOSS Quasar) & $1.526$ & $0.342 \pm 0.070$ & \cite{eBOSS:2018yfg}\\
SDSS-IV (eBOSS Quasar) & $1.944$ & $0.364 \pm 0.106$ & \cite{eBOSS:2018yfg}\\ 
\multicolumn{4}{c}{\vspace{-8pt}} \\
\hline
\end{tabular} 
\end{table} 

We list the full set of $\hf$ measurements used in this work in Table~\ref{tab:fhat_val}. This data set, containing 22 values between $z=0.001$ and $z=1.944$, is based on the Gold-2017 compilation of robust and independent measurements~\cite{Nesseris:2017vor}, being updated with more recently released data from SDSS-IV~\cite{eBOSS:2020yzd, eBOSS:2018yfg} as well as a very low redshift value from the 2MTF survey~\cite{Howlett:2017asq}. From Eq.~\eqref{eq:EG}, we see that to obtain $E_G$ at redshift $z$, measurements of $\hf$ and of $\hJ$ at the same redshift are needed. Since $\hf$ is measured from spectroscopic galaxy surveys, while $\hJ$ is obtained from lensing surveys, the redshifts typically differ. It is therefore necessary to propagate the measurements of $\hf$ to the desired redshifts, where measurements of $\hat{J}(z)$ are available. More precisely, we aim to reconstruct $\hf$ at the four effective redshifts of the DES MagLim lens sample, where measurements of $\hJ$ have been obtained in Ref.~\cite{Tutusaus:2023aux} as summarised in the first two columns of Table~\ref{tab:Jhat_fhat_EG}. Note that in principle we could instead reconstruct the Weyl evolution, $\hJ$, at the redshifts where we have measurements of $\hf$. This would however lead to less precise measurements of $E_G$ since the Weyl evolution is only measured at four redshifts, while $\hf$ is measured at 22 redshifts, allowing for a better reconstruction.

A model-independent method to reconstruct $\hf$ is provided by Gaussian processes. However, as shown in Ref.~\cite{Perenon:2021uom}, current data are not precise enough to allow for a robust reconstruction with this method, as the results are dominated by the prior range on the kernel hyperparameters. Therefore, we instead use spline interpolation, treating the values of $\hf$ at the four DES MagLim effective redshifts as four free parameters.  
We interpolate between these free parameters using cubic splines and minimise the difference between the interpolated curve and the measurements of $\hf$. We have examined choices for the knots in the spline interpolation other than the MagLim effective redshifts, such as linearly spaced values along the redshift range of the measured data, which affects the reconstruction only marginally (at sub-percent level for the mean values and percent level for the errors at the MagLim redshifts). Moreover, we have verified that four redshift knots provide the ideal number of parameters following the Akaike information criterion~\cite{Akaike:1974vps}. 

\begin{table}
\centering
\caption{We list the first four effective redshifts of the DES MagLim sample along with the respective values of $\hat{J}(z)$ obtained in Ref.~\cite{Tutusaus:2023aux} (using CMB priors and standard scale cuts), and the values of $\hat{f}(z)$ and $E_G(z)$ obtained in this work.} \label{tab:Jhat_fhat_EG}
\begin{tabular}{c c c c} 
\hline 
\multicolumn{4}{c}{\vspace{-8pt}} \\
$z$ & $\hat{J}(z)$ & $\hat{f}(z)$ & $E_G(z)$ \\
\multicolumn{4}{c}{\vspace{-8pt}} \\
\hline\hline 
\multicolumn{4}{c}{\vspace{-8pt}} \\
$0.295$ & $0.325 \pm 0.015$ & $0.459 \pm 0.019$ & $0.447 \pm 0.027$ \\
$0.467$ & $0.333\pm 0.018$ & $0.467 \pm 0.020$ & $0.378 \pm 0.026$ \\
$0.626$ &  $0.387\pm 0.027$ & $0.461 \pm 0.021$ & $0.396 \pm 0.033$ \\
$0.771$ & $0.354\pm 0.035$  & $0.448 \pm 0.024$ & $0.345 \pm 0.039$\\
\multicolumn{4}{c}{\vspace{-8pt}} \\
\hline
\end{tabular} 
\end{table}

\section{Results}
\label{sec:results}

\begin{figure}
    
    \centering
    \includegraphics[width=0.45\textwidth]{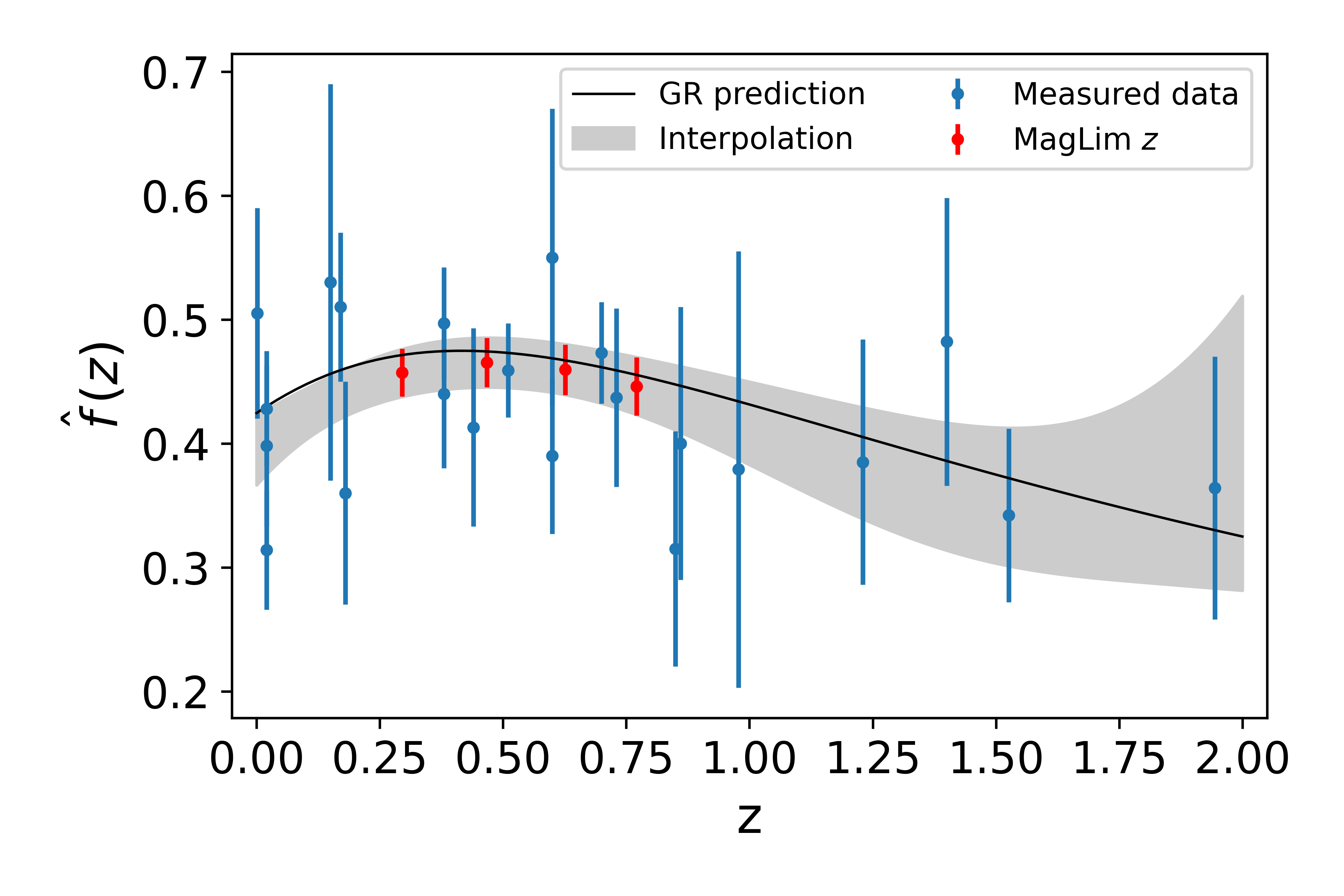} 
    \caption{Literature values of $\hat{f}(z)$ (blue points) along with our reconstruction at the DES MagLim effective redshifts (red points). In grey, we show the $1\sigma$ error bands of the spline reconstruction, while the black line shows the prediction assuming Planck 2018 cosmological parameters~\cite{Planck:2018vyg} and no deviations from GR.}
    \label{fig:fhat_interpolation}

\end{figure}

In Fig.~\ref{fig:fhat_interpolation}, we show in blue the 22 measurements of $\hf$ summarised in Table~\ref{tab:fhat_val}, and in red we show the values and error bars inferred at the four DES redshifts, which we list in the third column of Table~\ref{tab:Jhat_fhat_EG}. We note that, similarly to Refs.~\cite{Sagredo:2018ahx, Perenon:2019dpc}, we take the covariance matrix of the eBOSS Quasar and WiggleZ measurements into account. All remaining measurements of $\hf$ have negligible correlations~\cite{eBOSS:2020yzd, Nesseris:2017vor}. 
We see that $\hf$ is well reconstructed with precision between $4.3\%$ and $5.4\%$. This is not surprising, as we are using the knowledge of $\hf$ measured at 22 redshifts between 0.001 and 1.944, to obtain the values of $\hf$ at four new redshifts. As such, each red data point combines information from the whole set of measurements. We also note that our reconstruction is, within $1\sigma$, well-compatible with the GR prediction (black line) along the whole redshift range.

\begin{figure}    \includegraphics[width=0.45\textwidth]{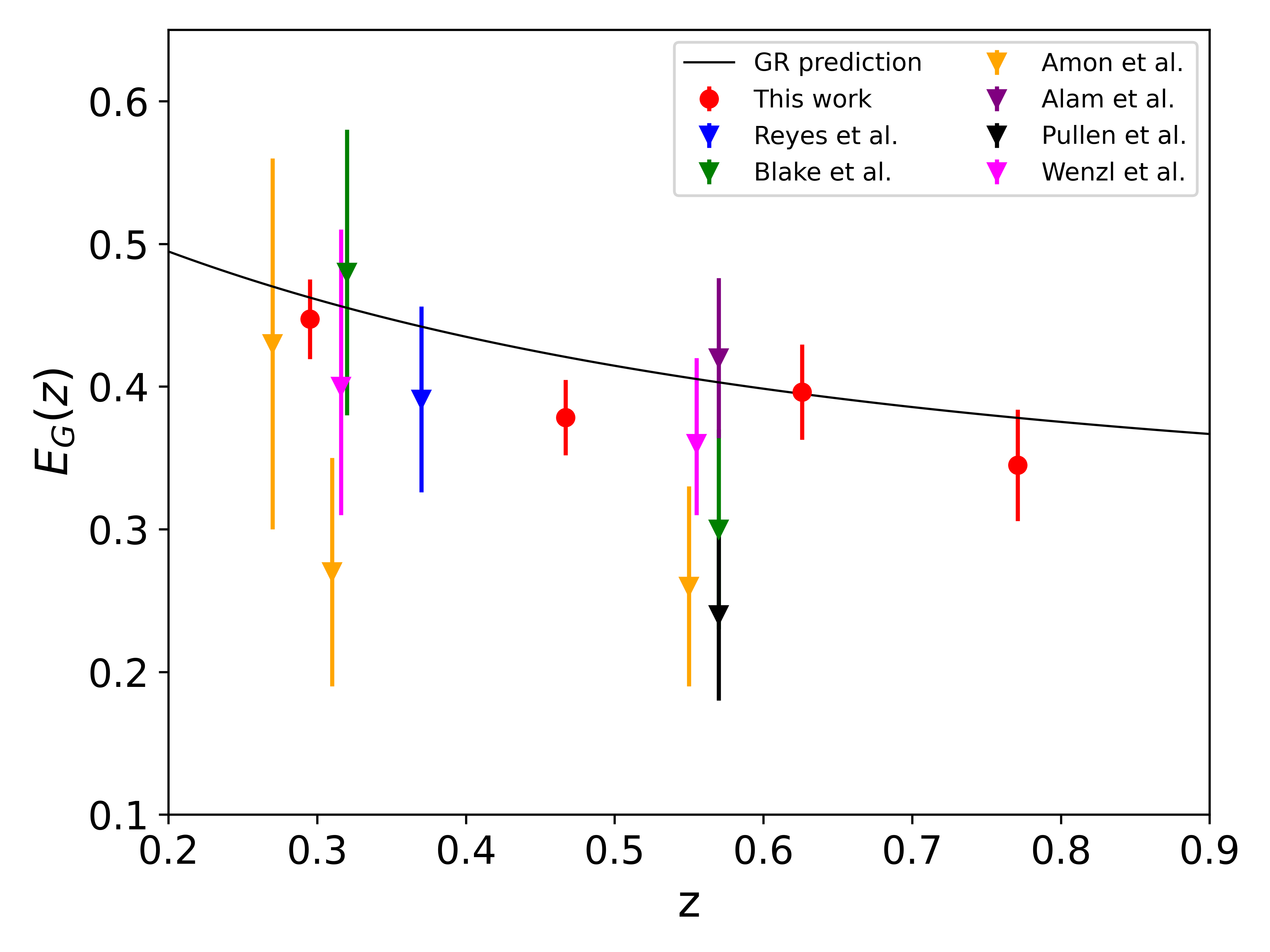}
    \caption{We show our measurements of $E_G$ (red) compared to various literature values.}
    \label{fig:EG_results}
\end{figure}

In Fig.~\ref{fig:EG_results}, we show in red the values of $E_G$ that we obtain with our method. We note that, to obtain the error bars, we have taken the full covariance of the $\hat{J}$ measurement from Ref.~\cite{Tutusaus:2023aux} into account, as well as the full covariance of the four $\hat{f}$ values obtained with our interpolation method. For comparison, we also show values previously obtained in literature (summarised in Table~\ref{tab:EG}), either using cosmic shear~\cite{Alam:2016qcl, Amon2018MNRAS.479.3422A, Reyes:2010tr, Blake:2016fcm} or CMB lensing~\cite{Pullen2016, Wenzl:2024sug}. As we see from Eq.~\eqref{eq:EG}, to obtain $E_G$ values that can be compared with the literature, we need to normalise the ratio $\hJ/\hf$ by the Hubble parameter. We assume here that the evolution of $\HH(z)$ obeys $\Lambda$CDM predictions with cosmological parameters given by Planck~\cite{Planck:2018vyg}. Note that this is not a limitation of our methodology: we could easily measure directly the ratio $\hJ/\hf$ to test for deviations from GR instead of using Eq.~\eqref{eq:EG}. The normalisation is only needed to compare with previous measurements of $E_G$. 

\begin{table}
\centering
\caption{List of literature values of $E_G$.} \label{tab:EG}
\begin{tabular}{c c c}
\hline
\multicolumn{3}{c}{\vspace{-8pt}} \\
$z$ & $E_G$ & Ref. \\
\multicolumn{3}{c}{\vspace{-8pt}} \\
\hline \hline 
\multicolumn{3}{c}{\vspace{-8pt}} \\
0.37 & $0.392 \pm 0.065$ & \cite{Reyes:2010tr} \\
0.32 & $0.48 \pm 0.10$ & \cite{Blake:2016fcm} \\
0.57 & $0.30 \pm 0.07$ & \cite{Blake:2016fcm} \\
0.27 & $0.43 \pm 0.13$  & \cite{Amon2018MNRAS.479.3422A} \\
0.31 & $0.27 \pm 0.08$ &\cite{Amon2018MNRAS.479.3422A} \\
0.55 & $0.26 \pm 0.07$ & \cite{Amon2018MNRAS.479.3422A} \\
0.57 & $0.420 \pm 0.056$ & \cite{Alam:2016qcl} \\
0.57 & $0.24 \pm 0.06$ & \cite{Pullen2016} \\
\multicolumn{3}{c}{\vspace{-10pt}} \\
 0.316 & $0.40^{+0.11}_{-0.09}$ & \cite{Wenzl:2024sug} \\
 \multicolumn{3}{c}{\vspace{-8pt}} \\
  0.555 & $0.36^{+0.06}_{-0.05}$ & \cite{Wenzl:2024sug} \\
 \multicolumn{3}{c}{\vspace{-8pt}} \\
 \hline
\end{tabular}
\end{table}

We see that the values that we obtain in the first, third and fourth redshift bin are in perfect agreement with the GR prediction within $1\sigma$,   shown by the black line in Fig.~\ref{fig:EG_results}. In the second bin, we find that our value is slightly below the GR prediction, at $1.6\sigma$. This behaviour can be traced back to the value of $\hJ$ in that bin, which is $2.8\sigma$ below the GR value~\cite{Tutusaus:2023aux}. Note that, in the first redshift bin, $\hJ$ is also below the GR value (by 2.3$\sigma$). However, dividing by $\hf$ with a best-fit value slightly below GR leads to a measurement of $E_G$ in perfect agreement with GR. This shows that, while $E_G$ provides an interesting consistency check of GR, it may be insensitive to some specific types of modifications of gravity. More specifically, models that lead to a growth rate of structure and a Weyl potential both lower or higher than in GR may not impact $E_G$ significantly.

We summarise our results for $E_G$ in the last column of Table~\ref{tab:Jhat_fhat_EG}, and we find that $E_G$ is measured with a precision between $6.0\%$ and $11.3\%$, compared to $13.3-30.2\%$ precision for the literature values in Table~\ref{tab:EG}. On one hand, this is due to the fact that our method does not rely on the correlation of cosmic shear with the same spectroscopic sample that is used to measure $\hf$. 
This is contrary to common methods  to obtain $E_G$, where the use of the same sample for these two measurements is crucial in order to cancel the impact of the galaxy bias~\cite{Zhang:2007nk}.
With our method, instead, we measure directly the Weyl evolution $\hJ$, which does not depend on the bias. Naturally, the Weyl evolution is in fact degenerated with the bias in the galaxy-galaxy lensing signal. However, as shown in Ref.~\cite{Tutusaus:2023aux}, by combining galaxy-galaxy lensing with galaxy clustering (i.e., the clustering of the lenses), both $\hJ$ and the galaxy bias can be measured. Our method allows us therefore to use the full statistical power of the photometric DES MagLim sample, containing 7.6 million galaxies in the first four bins~\cite{DES:2021bpo}, which is $10-100$ times more than the spectroscopic samples used for previous $E_G$ measurements~\cite{Reyes:2010tr, Blake:2016fcm, Amon2018MNRAS.479.3422A, Alam:2016qcl, Pullen2016, Wenzl:2024sug}.

The second reason for the high level of precision of our results for $E_G$ is due to the fact that we combine measurements of $\hf$ over a wide range of redshift, to infer $\hf$ at the four redshifts where $\hJ$ measurements are available. We note that, since previous measurements have employed information from surveys such as BOSS and GAMA, we expect some mild correlations between our measurements and previous measurements of $E_G$. However, we employ a much larger wealth of data for $f\sigma_8$, leading to more precise measurements of $E_G$. 

Finally, we emphasise that our method has the advantage of taking into account contaminations such as magnification bias~\cite{Ghosh:2018nsm} and redshift-space distortions in the galaxy-galaxy lensing and photometric galaxy clustering correlations. As explained in Ref.~\cite{Tutusaus:2023aux} and following the DES analysis~\cite{DES:2021bpo}, magnification bias and redshift-space distortions are consistently included in the modelling. Since these effects are subdominant, it is enough to model them in GR to obtain robust measurements of $\hJ$ (changing the modelling of these effects only impacts $\hJ$ very mildly). On the contrary, as shown in Ref.~\cite{MoradinezhadDizgah:2016pqy}, standard methods are affected by lensing magnification, which may spoil the bias-independence of $E_G$.

\section{Conclusion}

In this letter, we have presented new measurements of $E_G$ at four different redshifts. These measurements are probing the ratio between the Weyl potential and the motion of galaxies. They therefore provide a direct and robust test of the consistency of GR, which links the motion of astrophysical objects to the distorted geometry of the Universe. We have found that  $E_G$ is  consistent with GR predictions at the $1\sigma$ level at three of the four bins, and at $1.6\sigma$ level at the second bin. Our measurements are significantly more precise than previous measurements, providing more stringent constraints on deviations from GR. 

Interestingly, previous measurements of $E_G$ have been inconclusive regarding the compatibility with GR, in particular at redshifts between the second and third bin of our analysis. While Refs.~\cite{Alam:2016qcl, Wenzl:2024sug} show agreement with GR, Refs.~\cite{Amon2018MNRAS.479.3422A, Pullen2016, Blake:2016fcm} show mild tensions reaching up to $2.6\sigma$. Our method will be vital to further investigate the evolution of $E_G$ with redshift, once more measurements of $\hJ$, covering especially the range between the second and third bin, will be available from future surveys.

\begin{acknowledgments}

We thank Maria Berti for useful discussions. C.B.~and N.G.~acknowledge support from the European Research Council (ERC) under the European Union's Horizon 2020 research and innovation program (grant agreement No.~863929; project title ``Testing the law of gravity with novel large-scale structure observables"). C.B.~acknowledges support from the Swiss National Science Foundation.

\end{acknowledgments}



\bibliography{EG_withJ}
\bibliographystyle{apsrev4-1}

\end{document}